\documentclass{PoS}

\usepackage{graphicx}
\usepackage{natbib}
\newcommand{\HI}{H{\,\small I}}
\newcommand{\ltsima} {$\; \buildrel < \over \sim \;$}
\newcommand{\gtsima} {$\; \buildrel > \over \sim \;$}
\newcommand{\lta} {\lower.5ex\hbox{\ltsima}}
\newcommand{\gta} {\lower.5ex\hbox{\gtsima}}

\newcommand{\kms}{km\ s$^{-1}$}
\newcommand{\FRI}{FR{-\small I}}
\newcommand{\FRII}{FR{-\small II}}
\title{Cold gas $\&$ mergers: fundamental difference in HI properties of different types of radio galaxies?}

\ShortTitle{Large-scale HI in radio galaxies}

\author{\speaker{Bjorn Emonts}\thanks{Funded by the Netherlands Organisation for Scientific Research (NWO) under Rubicon grant 680.50.0508}\\
        Columbia University, Department of Astronomy, Mail Code 5246, 550 West 120th Street, New York, N.Y. 10027, USA\\
        E-mail: \email{emonts@astro.columbia.edu}}

\author{Raffaella Morganti, Tom Oosterloo\\
        Netherlands Foundation for Research in Astronomy (ASTRON), Postbus 2, 7990 AA Dwingeloo, the Netherlands\\
        E-mail: \email{morganti@astron.nl,oosterloo@astron.nl}}

\author{Jacqueline van Gorkom\\
        Columbia University, Department of Astronomy, Mail Code 5246, 550 West 120th Street, New York, N.Y. 10027, USA\\
        E-mail: \email{jvangork@astro.columbia.edu}}

\abstract{We present results of a study of large-scale neutral hydrogen (\HI) gas in nearby radio galaxies. We find that the early-type host galaxies of different types of radio sources (compact, \FRI\ and \FRII) appear to contain fundamentally different large-scale \HI\ properties: enormous regular rotating disks and rings are present around the host galaxies of a significant fraction of low power compact radio sources, while no large-scale \HI\ is detected in low power, edge-darkened \FRI\ radio galaxies. Preliminary results of a study of nearby powerful, edge-brightened \FRII\ radio galaxies show that these systems generally contain significant amounts of large-scale \HI, often distributed in tail- or bridge-like structures, indicative of a recent galaxy merger or collision. Our results suggest that different types of radio galaxies may have a different formation history, which could be related to a difference in the triggering mechanism of the radio source. If confirmed by larger studies with the next generation radio telescopes, this would be in agreement with previous optical studies that suggest that powerful \FRII\ radio sources are likely triggered by galaxy mergers and collisions, while the lower power \FRI\ sources are fed in other ways (e.g. through the accretion of hot IGM). The giant \HI\ disks/rings associated with some compact sources could - at least in some cases - be the relics of much more advanced mergers.}

\FullConference{From planets to dark energy: the modern radio universe\\
		 October 1-5 2007\\
		 University of Manchester, Manchester, UK}

\begin{document}

\section{Introduction}
Radio sources are generally hosted by early-type galaxies. A significant fraction of these radio galaxies, in particular the more powerful ones, show optical peculiarities (tail, bridges, shells, etc.) that are indicative of galaxy mergers or interactions \citep{hec86,bau92}. This has led to the suggestion that galaxy mergers or collisions could be the trigger for the radio-AGN activity in these systems. 

Here we present the large-scale neutral hydrogen (\HI) properties of different types of nearby radio galaxies in order to investigate whether these systems have been involved in a recent or past gas-rich merger or interaction. Simulation show that in a major merger between gas-rich galaxies, large-scale gaseous tidal structures can be expelled from the merging system, which after several rotations ($\gtrsim$ Gyr) can partially fall back onto the host galaxy and settle in a low-density, regular rotating disk or ring \citep{bar02}. \HI\ observations -- in particular when combined with a stellar population analysis of the host galaxy \citep{tad05} -- therefore provide an excellent tool to trace and date galaxy mergers on relatively long time-scales, which can be compared with the age of the radio-loud AGN in these galaxies in order to investigate a possible connection with the triggering of the radio source.

\section{HI in lower power radio galaxies (compact $\&$ FR-I)}

\begin{figure*}[t]
\centering
\includegraphics[width=12.5cm]{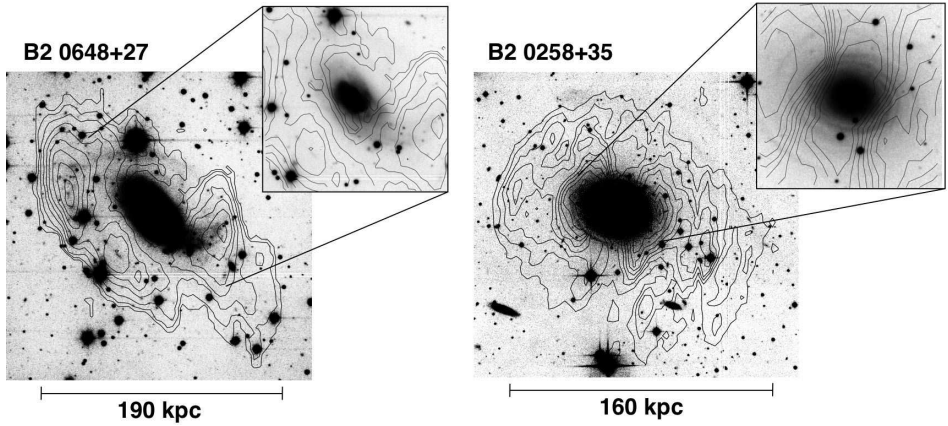}
\caption{Two HI-rich radio galaxies from our B2 sample. {\sl Left:} B2~0648+27 \citep{emo06} -- the HI ring (contours) in this advanced merger system follows a faint stellar tail seen in deep optical B+V band imaging (gray-scale). Contours HI: from 0.22 to 2.1 $\times 10^{20}$ cm$^{-2}$. {\sl Right:} B2~0258+35 -- the host galaxy is classified as early-type, but deep optical B-band imaging (gray-scale) reveals faint spiral structure in the disk. Contours HI: from 0.34 to 3.0 in steps of 0.33 $\times 10^{20}$ cm$^{-2}$. See Table 1 for more details.}
\label{fig:B2detections}
\end{figure*}

Our first study comprises a complete sample of all non-cluster radio galaxies from the B2-catalog ($F_{\rm 408 MHz} \gtrsim 0.2$ Jy) up to a redshift of $z \sim 0.04$. This sample contains low power compact radio sources as well as sub-relativistic, edge-darkened Fanaroff $\&$ Riley \citep{fan74} type-I (\FRI) sources. We detect large-scale \HI\ in emission in six ($\sim$25$\%$) of our sample sources. Two of these are shown in Fig.~\ref{fig:B2detections} (images of the other \HI\ detections can be found in Emonts et al. 2007 \citep{emo07}). Details of the \HI\ structures are given in Table \ref{tab:properties}. The two most striking results from this low power sample are:
\begin{itemize}
\item{Large-scale \HI\ structures are mainly found around the host galaxies of several compact radio sources, while for none of the extended ($>20\ \rm kpc$) \FRI\ sources \HI\ is detected outside the optical body of the host galaxy (see also Fig.~\ref{fig:HImassplot});}
\item{These detected \HI\ structures are mostly large-scale regular rotating disks and rings with diameters up to 190 kpc and \HI\ masses up to $2 \times 10^{10} M_{\odot}$!}
\end{itemize}
In order to investigate the nature of these large-scale \HI\ disks/rings in more detail, we studied one of the radio galaxies, B2~0647+27, in great detail. As is published in Emonts et al. 2006 \citep{emo06}, a detailed spectroscopic stellar population analysis of this system shows that a 0.3 Gyr post-starburst stellar population is present throughout this system, which could have given B2~0648+27 the appearance of an (Ultra-) Luminous Infra-Red Galaxy in the first epoch after the starburst was triggered. New deep optical imaging of this system (see Fig.~\ref{fig:B2detections} {\sl - left}) shows that a faint stellar tail is present in the \HI\ ring. The morphology of the \HI\ ring, the faint stellar tail and the presence of a post-starburst stellar population in B2~0648+27 suggest that this galaxy is an {\sl advanced} major merger. The merger event in this system must have occurred roughly 1.5 Gyr ago, after which \HI\ and stellar tails -- which were expelled during the merger -- had time to fall back onto the host galaxy and settle in the rotating ring-like structure that we observe. B2~0648+27 is an excellent example of a galaxy in which the different stages of merger, starburst and AGN activity can be traced (see Emonts et al. 2006 \citep{emo06} for more details). 

We note, however, that the formation mechanism of the large-scale \HI\ disks in the other B2 radio galaxies remains uncertain. Figure \ref{fig:B2detections} {\sl (right)} shows, for example, that the \HI\ disk B2~0258+35 has a faint and tightly wound spiral structure in deep optical imaging. This could be the result of a much older merger event, but perhaps it indicates that B2~0258+35 is rather a more typical spiral-type galaxy. Further studies are necessary to determine to nature of this and the other \HI\ disks. 


\begin{table}[b]
\small
\centering
\caption{Properties of the \HI-detected radio galaxies}
\label{tab:properties}
\begin{tabular}{llllcccl}
\hline
Host & Ref. &\multicolumn{2}{l}{Source type} & {\sl M}$_{\rm HI}$ & {\sl D}$_{\rm HI}$ & Morph. & Remarks \\
Galaxy & &\multicolumn{2}{l}{(+ diam./kpc)} & ({\sl M}$_{\odot}$) & (kpc) & HI &  \\
\hline
\hline
B2~0258+35 & \citep{emo07} & C &(1.4)    &  2$\times$10$^{10}$ & 160 & disk & faint optical disk \\ 
B2~0648+27 & \citep{emo06} & C &(1.3)    &  9$\times$10$^{9}$  & 190 & ring & advanced merger \\
B2~0722+30 & \citep{emo07} & FR-I &(14)   &  2$\times$10$^{8}$  & 15  & disk & spiral host galaxy \\
B2~1217+29 & \citep{mor06} & C &($>$1)   &  7$\times$10$^{8}$  & 37  & disk & classical elliptical (NGC~4278)\\
B2~1322+36 & \citep{emo07} & FR-I &(19)  & 7$\times$10$^{7}$   & 20  & blob & + resolved \HI\ absorption \\
NGC~3894   & \citep{emo07} & C &(1.6)    &  2$\times$10$^{9}$  & 105 & ring & faint optical dust-lane \\
NGC~612    & \citep{emo08} & FR-II &(473)& 2$\times$10$^{9}$   & 140 & disk+bridge & S0 with star-forming disk\\
3C~305$^{*}$     &     & FR-II &(11) &  $6 \times$10$^{8}$& 106 & tail & + prominent central absorption \\
3C~382$^{*}$     &     & FR-II &(284)&  $2 \times$10$^{9}$& 92& bridge/tail & \HI\ maybe part of environment? \\
3C~192$^{*}$     &     & FR-II &(331)&  3$\times$10$^{8}$  & 110 & bridge& \HI\ bridge detected in absorption \\
3C~390.3$^{*}$   &     & FR-II &(349)&  2$\times$10$^{9}$  & 172 & tail  & absorption; maybe environment? \\
\hline
\multicolumn{6}{l}{* preliminary results -- {\sl need to be verified}} & \multicolumn{2}{l}{$\ \ \ \ \ \ \ \ \ \ \ \ \ \ \ \ \ \ $ Assumed $H_{0} = 71$ \kms\ Mpc$^{-1}$} \\
\end{tabular}\\
\end{table}

\section{HI in higher power radio galaxies (FR-II)}

In order to investigate the large-scale \HI\ properties of higher power, edge-brightened \FRII\ radio galaxies (with relativistic radio jet ending in a hot-spot), we observed a sample of the nearest ($z \lesssim 0.06$) powerful radio galaxies. Because the volume density of \FRII\ sources in the nearby Universe is extremely low, \HI\ can be mapped only for a very limited number of \FRII\ radio galaxies with the sensitivity of current-day radio telescopes (and even then these \FRII\ sources are not as powerful as their high-$z$ counterparts).

\begin{figure*}[t!]
\centering
\includegraphics[width=14cm]{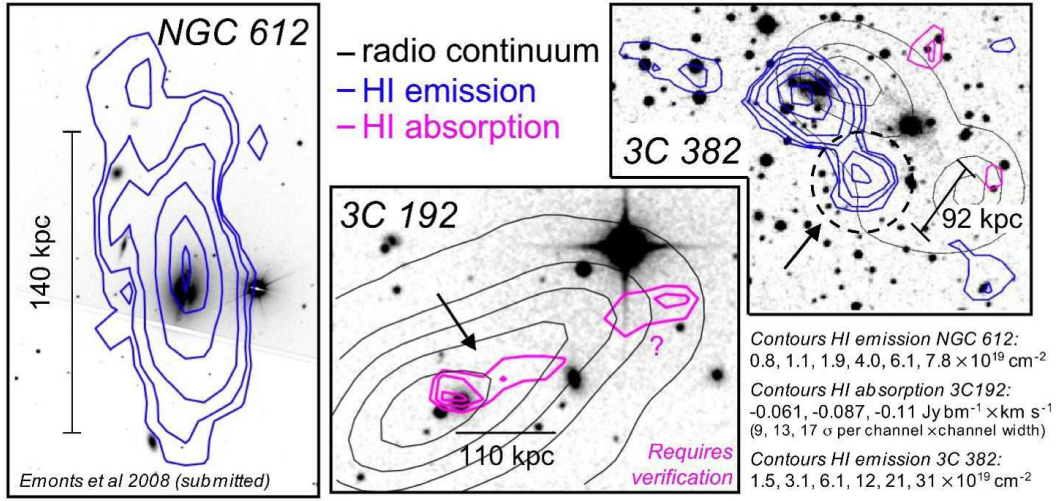}
\caption{Preliminary total intensity maps of large-scale \HI\ detected around three nearby \FRII\ radio galaxies. The arrows point to the \HI\ structures that we assign to the radio galaxies (see Table 1 for details). The different contours are indicated in the figure (for clarification, the extended radio source in NGC~612 is omitted from the plot).}
\label{fig:HIFRII}
\end{figure*}

Preliminary results show that 5 out of 7 \FRII\ radio galaxies that we studied so far appear to contain large-scale \HI\ (three examples are shown in Fig.~\ref{fig:HIFRII}, while Table \ref{tab:properties} summarizes the preliminary results of the remaining detections). This \HI\ detection rate is significantly higher than the detection rate for the lower power \FRI\ radio galaxies. NGC~612 (Fig.~\ref{fig:HIFRII} {\sl - left}) contains a large-scale rotating disk \citep{emo08}, very similar in morphology to the disks and rings detected around some compact sources from our B2-sample. The \HI\ features detected around the other \FRII\ radio galaxies resemble more irregular tail- or bridge-like structures. For 3C~192 (Fig.~\ref{fig:HIFRII} {\sl - middle}) a large-scale \HI\ bridge is detected in absorption against the strong radio continuum (although the signal is weak and only detected at the 4$\sigma$ level in the individual channel maps). We note that for 3C~382 (Fig.~\ref{fig:HIFRII} {\sl - right}) and 3C~390.3 (Table \ref{tab:properties}), it is uncertain whether the \HI\ is directly related to the radio galaxy or part of its \HI-rich environment. {\sl Our preliminary \HI\ detections need to be verified} and additional observations are planned in order to accurately map the large-scale \HI\ structures around these \FRII\ radio galaxies.

\section{Discussion}
\label{sec:discussion}

\begin{figure}[t]
\centering
\includegraphics[width=10cm]{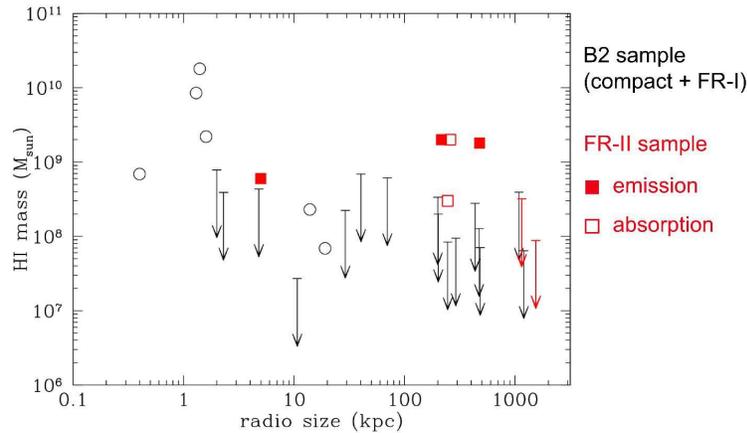}
\caption{Large-scale \HI\ mass plotted against the total linear extent of the radio source. Black circles and arrows represent the sources from our B2 sample (compact + \FRI), while red squares and arrows are preliminary results from our \FRII\ sample. In case of non-detection a conservative upper limit (3$\sigma$ across 400 \kms) was estimated.}
\label{fig:HImassplot}
\end{figure}

Our results of large-scale \HI\ around nearby radio galaxies imply that there could be {\sl a fundamental difference in large-scale \HI\ content between the various types of radio galaxies}, which is visualized in more detail in Fig.~\ref{fig:HImassplot}. While the host galaxies of several low power compact radio sources contain a large-scale regular rotating \HI\ disk or ring (possibly as the result of an advanced merger), the extended \FRI\ radio galaxies in the same sample lack any large-scale \HI\ structures outside the optical body of the host galaxy up to a conservative detection limit of a few $\times 10^{8} M_{\odot}$. On the other hand, preliminary results for a small sample of the more powerful \FRII\ radio sources show that most of their host galaxies do contain significant amounts of \HI, often distributed in tail- or bridge-like structures. If confirmed by future studies with greater statistical significance and lower detection limits, this could mean that there is a fundamental difference in the formation history of different types of radio galaxies and, related, the triggering of their radio source. This would be in agreement with optical studies that suggest that powerful \FRII\ sources are generally fueled by major mergers, while \FRI\ source are likely powered in another way, for example through the accretion of gas from the hot inter-galactic medium \citep{hec86,bau92}.

Future instruments, with eventually the Square Kilometre Array, will be essential to verify our results by observing \HI\ emission in large statistical samples of radio galaxies.

\end{document}